\documentclass[prl,aps,twocolumn,showpacs,bibnotes,epsf]{revtex4}
\usepackage{graphicx, verbatim}
\newcommand{\Journal}[4]{#1 \textbf{#2}, #3 (#4)}
\usepackage{epstopdf}
\usepackage{CJK}
\usepackage[utf8]{inputenc}
\usepackage[english]{babel}
\usepackage{hyperref}
\hypersetup{colorlinks=true,linkcolor=blue,filecolor=magenta,urlcolor=cyan}

\begin{document}
\begin{CJK*}{GB}{gbsn}
\title{Control of spectral characteristics of spin-current auto-oscillator by electric field}
\author{R. H. Liu(刘荣华)$^{1,2}$}
\author{Lina Chen$^1$}
\author{S. Urazhdin$^2$}
\author{Y.W. Du(都有为)$^{1}$}
\affiliation{
$^1$National Laboratory of Solid State Microstructures, School of Physics and Collaborative Innovation \\Center of Advanced Microstructures, Nanjing University, Nanjing 210093, China\\
$^2$Department of Physics, Emory University, Atlanta, GA 30322, USA.}

\begin{abstract}
We study the effects of electrostatic gating on the magnetization auto-oscillations induced by the local injection of electric current into a ferromagnet/heavy metal bilayer. We find that the characteristic currents required for the excitation, the intensity and the spectral characteristics of the generated dynamical states can be tuned by the voltage applied to the metallic gate separated from the bilayer by a thin insulating layer. We show that the effect of electrostatic gating becomes enhanced in the strongly nonlinear oscillation regime at sufficiently large driving currents. Analysis shows that the observed effects are caused by a combination of electric field-dependent surface anisotropy and electric field-dependent contribution to the current-induced spin-orbit torques. The demonstrated ability to control the microwave emission and spectral characteristics provides an efficient approach to the development of electrically tunable microwave nano-oscillators.
\end{abstract}


\maketitle
\end{CJK*}
Significant progress has been made in recent years in the ability to control magnetism by magnetic field, current-induced spin torques, electrostatic and optical fields. All-electronic control of magnetization, achieved by the application of electric fields~\cite{matsukura}, electric currents~\cite{brataas}, or a combination of both, is particularly attractive, because it provides the benefits of high speed, low power, the possibility of downscaling for high-density applications, and is amenable to integration with the modern semiconductor electronics. Several approaches and mechanisms enabling all-electric control of magnetism have been extensively investigated in a number of materials and heterostructure geometries. For instance, the Curie temperature has been varied in magnetic semiconductors by the modulation of the carrier concentration~\cite{ohno,chiba}. Electrically driven magnetization reversal has been achieved in multiferroic materials due to its coupling to the electric polarization~\cite{lebeugle,heron,tokunaga} and in ultrathin ferromagnet/oxide heterostructures due to the voltage-dependence of the interfacial magnetic anisotropy~\cite{weisheit,maruyama,shiota}. Magnetization control by current, including current-induced magnetization reversal and auto-oscillations due to the spin transfer torque (STT), has been achieved in magnetic multilayer structures~\cite{slon1,berger,tsoiprl,cornellorig} and in heavy metal/ferromagnet heterostructures (FH)~\cite{miron,suzuki,liulq,demidov}. In FH bilayers, STT is produced by the pure spin current generated by a combination of the spin Hall effect (SHE) inside the heavy metal and the Rashba effect caused by the spin-orbit interaction (SOI) at the magnetic interface~\cite{diakonov,hirsch,rashba,manchon,pi}. The recently demonstrated electric field-assisted current-induced magnetization reversal in magnetic tunnel junctions has received a significant attention due to the flexibility of control and the high efficiency of the combined electric-field effect and current-induced spin torques~\cite{wang,liul}. Despite these promising features, the mechanisms of the electric field effect on the current-induced spin torques and magnetization dynamics are still being debated.

\begin{figure}[htbp]
%\vspace{+5mm}
\centering
\includegraphics[width=0.45\textwidth]{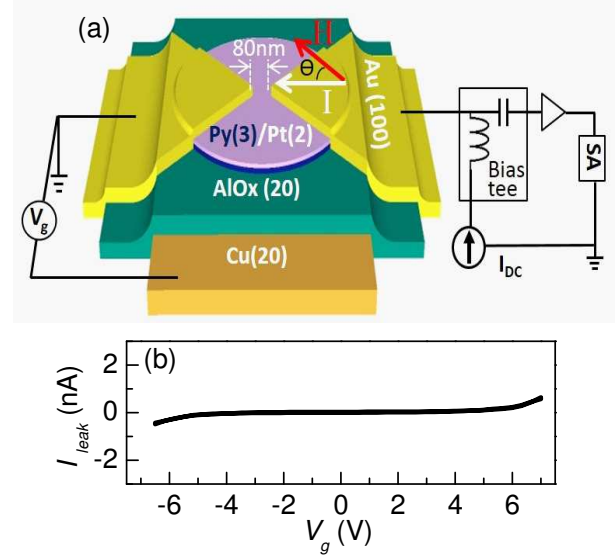}
\caption{(Color online). (a) Schematic of the device structure and the experimental setup. (b) Gate leakage current $I_{leak}$ {\it vs} gate voltage $V_g$ for a gated SCAO device.}\label{fig1}
%\vspace{+5mm}
\end{figure}

\begin{figure*}[!t]
\centering
\includegraphics[width=0.95\textwidth]{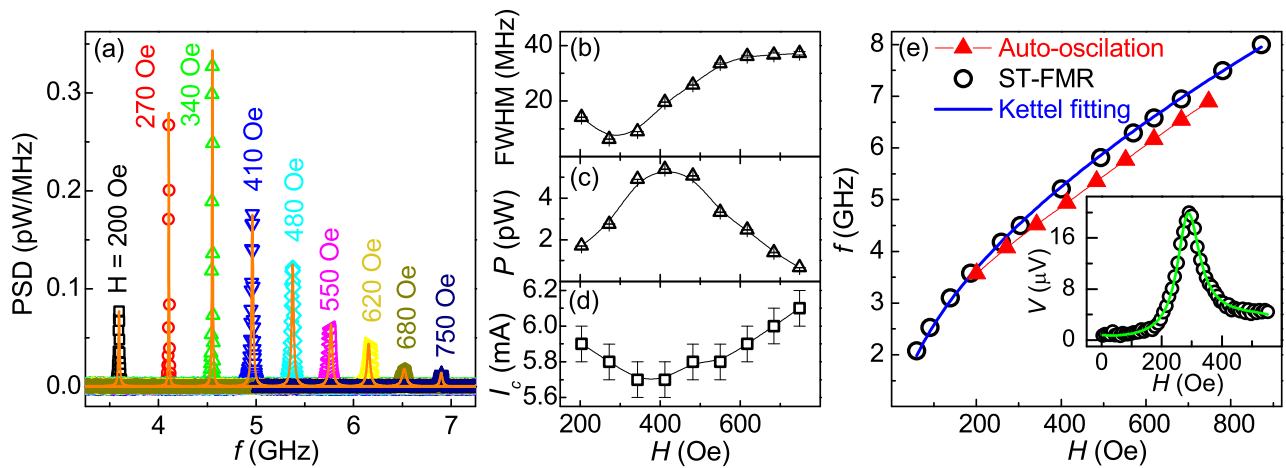}
\caption{(Color online). Dependence of the microwave generation characteristics on the magnetic field \emph{H}, at $V_g$ = 0. (a) Generation spectra (symbols) obtained at $I=6.7$~mA and the labeled values of the magnetic field. The curves are the results of fitting by the Lorentzian function. (b)-(d) Field dependencies of the minimum full width at half-maximum of the generation spectrum (b), the maximum integral intensity (c), and the critical current $I_c$ (d). (e) The spin torque FMR frequency (circles) and the maximum frequency of the auto-oscillation (triangles) {\it vs} \emph{H}. The solid curve is the result of fitting with the Kittel formula $f = \gamma\sqrt{H(H+H_d)}$, where $\gamma$ = 2.8 MHz/Oe is the gyromagnetic ratio, and the effective demagnetizing field $H_d$ has the best-fit value of $8380$~Oe determined by the magnetization of Py and the surface anisotropy. Inset: magnetic field dependence of the ST-FMR voltage obtained with a microwave current of $1$~mA rms at frequency $f_{ext}=4$~GHz. The curve is the best fit with a sum of a symmetric and an antisymmetric Lorentzian.} \label{fig2}
\end{figure*}

Here, we demonstrate the effects of electric field on the magnetization oscillation driven by the current-induced spin-orbit torques in a spin-current auto-oscillator (SCAO) based on the Permalloy(Py)/Pt bilayer. The highly coherent self-localized spin wave is excited in the extended Py film by the local injection of spin current, which is generated by a combination of spin Hall effect in Pt and the interfacial Rashba-type spin-orbit torque at the Pt/Py interface. We show that the electric field produced by the back-gating can modify the characteristic currents required for the auto-oscillation, as well as the spectral characteristics of the auto-oscillation, due to a combination of field-induced  spin-orbit torques and the modulation of interfacial magnetic anisotropy. We also demonstrate that the effects of electric field can be significantly enhanced in the strongly nonlinear dynamical regime of SCAO reached at sufficiently large driving currents. The demonstrated electrostatic gating effects provide an efficient approach to low-power control of the microwave generation characteristics of SCAO, enabling the development of new information processing technologies based on spin dynamics.

\begin{figure*}[!t]%[htbp]
\centering
\includegraphics[width=0.85\textwidth]{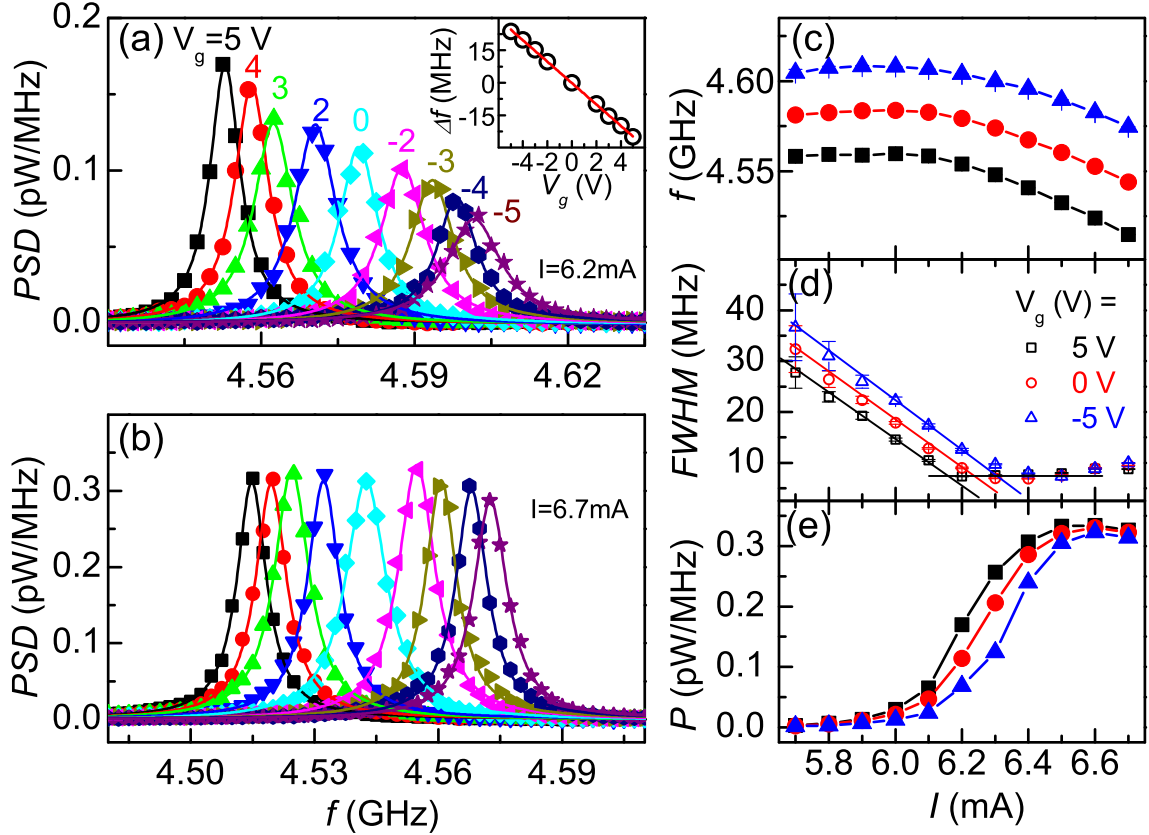}
\caption{(Color online).Effects of electrostatic gating on the microwave generation characteristics of SCAO. (a) Symbols: Power spectral density (PSD) of generation spectra at the labeled values of the gate voltage $V_g$ ranging from $\-5$~V to $5$ V, at $H = 340$~Oe and $I = 6.2$~mA. The curves are the results of fitting by the Lorentzian function. Inset: dependence of the central frequency shift $\Delta f(V_g) = f_0(V_g)- f_0(0)$ of the spectral peak on the gate voltage (symbols), and the linear fit of the data (line). (b) same as (a), at $I = 6.7$~mA. (c)-(e) Dependence of the central generation frequency $f_0$ (c), the linewidth FWHM (d), and the integral intensity $P$ (e) on current $I$ at $V_g$ = -$5$~V (triangles), $0$ (circles), and $5$~V (squares). The central frequency and the linewidth were determined by fitting the power spectra with the Lorentzian function, as shown in (a) and (b). The solid lines in (d) are the linear fits utilized to determine the effective excitation current shift, $\delta I(\pm 5~V) = \pm 0.07$~mA.}\label{fig3}
\end{figure*}

Figure.~\ref{fig1}(a) shows the schematic of the test device structure and the experimental setup. The device was fabricated on an annealed sapphire substrate by a combination of magnetron sputtering and e-beam lithography. The SCAO was based on a Py(3)/Pt(2) FH bilayer disk with a 4~$\mu$m diameter.  All thicknesses are given in nanometers. Two Au(100) electrodes with sharp tips separated by an $80$~nm gap, fabricated on top of the FH bilayer, were utilized to locally inject the electrical current in the bilayer. The spin Hall effect generated due to spin-orbit interaction in Pt by the current flowing through it produced a pure spin current flowing into the Py(3) layer, exerting STT on its magnetization. An additional torque was exerted on the Py magnetization due to the Rashba-like torque originating at the Py/Pt interface. This structure and its operation are similar to the planar point contact spin Hall nano-oscillators previously studied by optical~\cite{demidov} and magnetoelectronic techniques~\cite{liurh}. In our study, the device included an additional Cu(20) gate electrode fabricated underneath the FH bilayer, and electrically isolated from it by an AlO$_x$(20) layer. Figure~\ref{fig1}(b) shows the dependence of the leakage current $I_l$ between the FH and the gate electrode as a function of the voltage $V_g$ applied to the gate. The leakage does not exceed $100$~pA at gate voltages of up to $\pm5$~V, indicating a high quality of the AlO$_x$(20) insulator characterized by the breakdown electric field of more than $4$~MV/cm. All the measurements described below were performed at temperature $T=6$~K, and a fixed angle $\theta= 30^\circ$ between the in-plane field and the direction of current driving the SCAO.

Spectroscopic measurements of the microwave generation and the spin torque-induced ferromagnetic resonance (ST-FMR) performed at zero gate voltage allowed us to characterize the magnetic properties and the spectral characteristics of the SCAO [Fig.~\ref{fig2}]. Auto-oscillations, indicated by the abrupt emergence of a sharp spectral peak, were observed above the critical current $I_c\approx 5.7-6.1$~mA in the studied range of fields. The value of $I_c$ is significantly smaller than in the previously studied planar point contact spin Hall oscillators based on thicker Py/Pt bilayers~\cite{demidov,liurh}, due to a combination of the more efficient spin injection from the thin Pt layer~\cite{ulrichs} and the larger effects of spin current on the relatively thin Py layer.

Figure~\ref{fig2}(a) shows an example of the auto-oscillation spectra acquired at $I=6.7$~mA, and applied field ranging from $200$~Oe to $750$~Oe. The spectral peak was well approximated by the Lorentizian function, as shown by the solid curves. Spectral characteristics such as the full width at half maximum (FWHM), the central peak frequency $f_0$, and the generated power $P$ were extracted from the Lorentzian fitting. Fig.~\ref{fig2}(b) and (c) show the dependence of the minimum linewidth and the maximum integral intensity on field, both of which were observed at currents close to the same value $I_p$ corresponding to the largest amplitude of the microwave generation peak. The auto-oscillation frequency also exhibits a maximum at the same current value $I_p$. The linewidth is small at low fields, with a minimum of $6$~MHz at $H=270$~Oe, and gradually increases at larger fields to $36$~MHz at $H=750$~Oe [Fig.~\ref{fig2}(b)]. The integral intensity $P$ exhibits a non-monotonic dependence on field, with a broad maximum near the field $H\approx 400$~Oe somewhat larger than that corresponding to the minimum linewidth. The critical current $I_c$ for the onset of auto-oscillation exhibits a broad minimum of $5.7$~mA around the same field range [Fig.~\ref{fig2}(d)]. The observed variations of auto-oscillation characteristics are likely associated with the field dependence of the dynamical mode spectrum in the point contact region of the Py(3) film. In particular, the increase of the minimum linewidth and of the critical current, and the decrease of the maximum generated power at large fields, are indicative of the presence of a secondary dynamical mode whose frequency is close to the uniform FMR mode, resulting in decreased coherence and amplitude of  auto-oscillation~\cite{slavinprl,kim,nonlinear}.

The magnetic properties of the Py film and the nature of the auto-oscillation mode were established by utilizing the spin torque-driven ferromagnetic resonance (ST-FMR) technique~\cite{fuchs}. A microwave current with rms amplitude of $1$~mA was applied to SCAO, resulting in the magnetization oscillation at the same frequency. The oscillation amplitude was characterized by the dc voltage produced by mixing of the resistance oscillation with the applied current. The dependence of the dc voltage on field could be well fitted by a sum of a symmetric and an antisymmetric Lorentzian, as illustrated in the inset of Fig.\ref{fig2}(e) for a typical ST-FMR curve. Figure~\ref{fig2}e shows the field dependence of the  FMR frequency $f_{FMR}$ and of the maximum auto-oscillation frequency $f_0$ reached at $I=I_p$. These data show that the auto-oscillation frequency always remains lower than $f_{FMR}$, consistent with the properties of other in-plane magnetized SCAO~\cite{demidov,liurh}. These behaviors can be attributed to the nonlinear nature of the auto-oscillation mode, which forms a self-localized standing spin-wave "bullet"~\cite{slavinprl}. We note that the frequency of the auto-oscillation approaches $f_{FMR}$ at small $H$~200~Oe. This explains the low-field broadening of the auto-oscillation spectrum [Fig.~\ref{fig2}(b)], since a larger spectral overlap between the auto-oscillation and the linear spin wave spectrum results in larger radiation losses. The ST-FMR data can be fitted with the Kittel formula,\begin{equation}\label{Kittel}f = \gamma\sqrt{H(H+H_d)},\end{equation}, where $\gamma$ is the gyromagnetic ratio, and $H_d$ is the effective demagnetizing field. We note that spin-orbit coupling at Py/Pt and Py/AlO$_x$ interfaces results in uniaxial magnetic anisotropy of the ultrathin Py(3) layer, with the anisotropy axis normal to the film. This contribution can be taken into account as an additional term in the effective demagnetizing field $H_d=4\pi M - \frac{2K}{tM}$, where $K$ is the anisotropy coefficient, and $t$ = 3 nm is the thickness of the Py film. The magnitude $K = 0.27$ erg/cm$^{2}$ of the interfacial magnetic anisotropy was obtained from fitting the ST-FMR data using the saturation magnetization $M_s$ = 840 emu/cm$^{3}$ previously determined for similar Py films~\cite{liurh}.

Our central experimental result is the dependence of the magnetization oscillation characteristics on the electric field produced by the gate voltage $V_g$. We analyze this effect at field $H=340$~Oe, at which the SCAO produces a large generation power and a narrow spectral linewidth [Figs.~\ref{fig2}(b,c)]. Figures~\ref{fig3}(a) and (b) show the power spectral density of the oscillation spectra at $I=6.2$~mA and $6.7$~mA, respectively, for the bias voltage $V_g$ ranging from $-5$~V to $+5$ V. The shift $\Delta f(V_g) =f_0(V_g)-f_0(0)$ of the central oscillation frequency exhibits a linear dependence on $V_g$ with the slope of $4.6$ MHz/V at $I=6.2$~mA, as shown in the inset of Fig.~\ref{fig3}(a). In addition to this shift, at $I=6.2$~mA the intensity of the spectral peak also varies with the gate voltage. Meanwhile, at $I=6.7$~mA, gating produces only a frequency shift at a rate of $6$~MHz/V. The difference between the effects of gating at these two currents is closely correlated with the current dependence of the oscillation characteristics, as discussed below.

Figures~\ref{fig3}(c)-(e) show the current dependencies of the central oscillation frequency, the linewidth, and the generation power, measured at $V_g=-5$~V, $0$, and $5$~V. The frequency and the generation power slowly increase, and the linewidth linearly decreases at small currents. At currents above $~6.1$~mA, the generated power rapidly increases, the frequency exhibits a strong redshift, while the linewidth remains approximately constant. Similarly behaviors have been observed both in the spin valve nano-oscillators~\cite{urazhdin} and in spin Hall oscillators~\cite{liurh}, and are consistent with the nonlinear theory of magnetic nano-oscillators~\cite{nonlinear}. We note that the cross-over at $I=6.1$~mA superficially resembles a transition from the regime of spin current-enhanced thermal fluctuations at smaller currents to auto-oscillation at larger current. However, the oscillation frequency is below $f_{FMR}$ even at small currents, and therefore this dynamical state does not belong to the linear spin-wave spectrum and cannot be attributed to thermal fluctuations.

As is apparent from Fig.~\ref{fig3}(c), the effect of gating on the oscillation frequency can be described as mostly a vertical shift. This observation is consistent with the previous studies of the effects of gating in thin magnetic films, which demonstrated the possibility to modify the coercive field~\cite{weisheit} and even switch the magnetization direction~\cite{shiota}. These effects could be attributed to the modulation of the surface magnetic anisotropy, mostly due to the variation of the Fermi level at the interfaces~\cite{leix}. In our measurements, the frequency of the oscillation also depends on the interfacial  magnetic anisotropy, described as a contribution to the effective demagnetizing field $H_d$ in the Kittel formula Eq.~(\ref{Kittel}). Based on the data of Fig.~\ref{fig3}(c) and the Kittel formula, we estimate that the electrostatic gating modulates the interfacial anisotropy coefficient at the rate of $dK/dV_g=2.2\times 10^{-3}$~erg/(Vcm$^2$).

The effects of the electric field are not limited to the modulation of interfacial anisotropy. In particular, the oscillation linewidth can be modified by electric field at currents below the nonlinear crossover, where it linearly decreases with current [Fig.~\ref{fig3}(d)]. However, the linewidth is not significantly affected by gating at currents above the crossover. This effect can be approximately described as a horizontal shift, with the effective current modulation of $\delta I(V_g)=\pm 0.07$~mA for $V_g=\pm 5$~V. The effect of gating on the current-dependent generation power can be similarly described as an effective shift of the driving current [Fig.~\ref{fig3}(e)]. The main contribution to this shift likely comes from the electric-field dependent interfacial Rashba-like contribution to the spin-orbit torque, as demonstrated by the previous quasi-static measurements in similar FH heterostructures~\cite{liuprb}. This physical mechanism is expected to result in an effective rescaling of the driving current. For a small current range, rescaling is approximately equivalent to a shift.

\begin{figure}[htbp]
\centering
\includegraphics[width=0.4\textwidth]{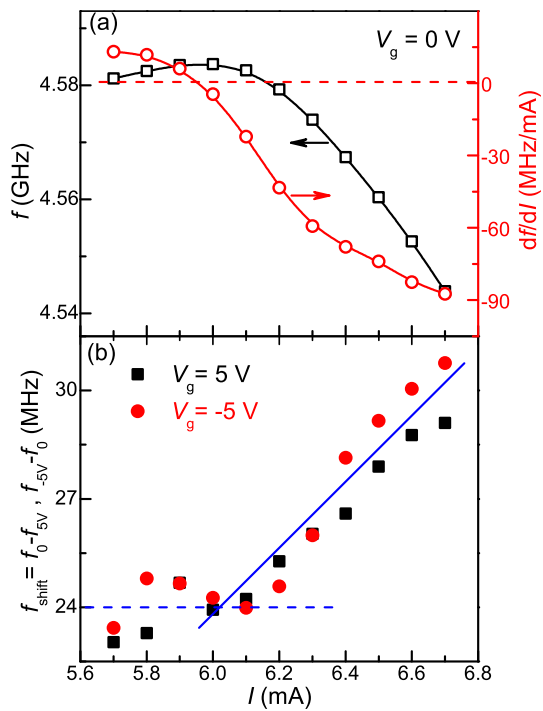}
\caption{(Color online).
Analysis of the current dependence of the electric field effects. (a) Center frequency of the auto-oscillation (left vertical axis) and its differential $df/dI$ (right vertical axis) {\it vs} $I$, at $V_g = 0$ and $H$ = 340 Oe. (b) The current dependence of the frequency shifts $-\Delta f(5V) = f(5V)-f(0)$ (squares) and $\Delta f(-5V)=f(-5V)-f(0)$ (circles), with the respective currents offset by $\Delta I=0.07$~mA for $V_g=5$~V and  $\Delta I(-5V)=-0.07$~mA for $V_g=-5$~V.
The dashed line indicates the current-independent frequency shift at $I < 6.1$~ mA. The solid line is a linear fit for the $I > 6.1$~mA data.}\label{fig4}
\end{figure}

The data of Fig.~\ref{fig3} clearly demonstrate two different effects of gating, one resulting from the field-dependent surface anisotropy, another from the Rashba-like torque. To quantitatively analyze their respective contributions to the oscillation frequency, we can write
\begin{equation}\label{gating}
df/dV_g = \frac{\partial f(I,K)}{\partial K} \frac{dK}{dV_g} + \frac{\partial f(I,K)}{\partial I} \frac{dI}{dV_g}
\end{equation}
where the first term represents the contribution of the surface anisotropy, and the second term comes from the spin-orbit torque. The second contribution is determined by the frequency nonlinearity of the oscillator described by $\partial f(I,K)/\partial I$. On the other hand, the voltage dependence of the interfacial anisotropy is not directly affected by the nonlinearity, and is expected to be approximately independent of the driving current.

The presented analysis demonstrates that the frequency nonlinearity of the oscillator can provide a significant contribution to the gating effects. In the studied SCAO, the magnitude of the nonlinearity remains small at currents up to $I=6.2$~mA, and exhibits a significant increase at larger currents, as illustrated by the current dependence of $f$ and $df/dI$ at $V_g=0$ in Fig.\ref{fig4}(a). Since the frequency nonlinearity is weak at $I<6.1$~mA, the contribution of the effective current shift to the frequency shift is small. As a consequence, gating effect is dominated by the variation of the surface anisotropy, independent of current. Indeed, the magnitude of the frequency shift $|\Delta f|\approx 24$~MHz at $V_g=\pm 5$~V remains approximately constant in this current range [Fig.~\ref{fig4}(b)].

At $I>6.1$~mA, the nonlinearity plays an increasingly significant role, as reflected by the increasing frequency shift in Fig.~\ref{fig4}(b).  We can estimate the value of $d I/d V_g\approx -0.014$~mA/V in the second term in Eq.~(\ref{gating}) from the dependencies of the linewidth and the generation power on the gate voltage [Fig.\ref{fig3}(d),(e)]. Based on the value of $\partial f/\partial I=-85$~MHz/mA at $I=6.7$~mA determined from Fig.~\ref{fig4}(a), we calculate the contribution $|\Delta f_I|=6.0$~MHz to the overall frequency shift at this current and $V_g=\pm 5$~V. This result is consistent with the total observed shift of $30$~MHz.

In summary, we have demonstrated that dynamical magnetization states excited by spin current in magnetic nano-oscillators based on spin-orbit torques can be  controlled by electrostatic gating. The observed effects are produced by two distinct contribution, one caused by the field-dependent interfacial magnetic anisotropy, another caused by the modulation of the Rashba-like interfacial torque. The former results mostly in the oscillation frequency shift, the latter is approximately equivalent to a shift of the driving current. Tuning of the oscillation characteristics by electrostatic gating represents an energy-efficient and fast alternative to the previously demonstrated tuning by current and variations of applied magnetic field~\cite{tsoiprl,cornellorig,urazhdin,rippard}.
Our results indicate that this mechanism can be particularly effective in strongly nonlinear oscillators. For instance, for a magnetic nano-oscillator with frequency nonlinearity $df/dI$ = 1.6 GHz/mA~\cite{rippard}, we estimate that the frequency tuning by gating can reach the efficiency of $22$~MHz/V at the base oscillation frequency of $24$~GHz, four times larger than $6$~MHz/V demonstrated in our device. Electric-field modulation of the current-induced magnetization oscillations and spin waves can be utilized for frequency mixing, synchronization, and in logic gates in spin wave-based electronic (magnonic) devices.

R.H.L, L.N.C and Y.W.D were supported by from National Key Research and Development Program of China (2016YFA0300803) and the Open Research Fund of Jiangsu Provincial Key Laboratory for Nanotechnology, Nanjing University. S.U. acknowledges support from NSF ECCS-1503878 and DMR-1504449.


\begin{thebibliography}{0}
\expandafter\ifx\csname natexlab\endcsname\relax\def\natexlab#1{#1}\fi
\expandafter\ifx\csname bibnamefont\endcsname\relax
  \def\bibnamefont#1{#1}\fi
\expandafter\ifx\csname bibfnamefont\endcsname\relax
  \def\bibfnamefont#1{#1}\fi
\expandafter\ifx\csname citenamefont\endcsname\relax
  \def\citenamefont#1{#1}\fi
\expandafter\ifx\csname url\endcsname\relax
  \def\url#1{\texttt{#1}}\fi
\expandafter\ifx\csname urlprefix\endcsname\relax\def\urlprefix{URL }\fi
\providecommand{\bibinfo}[2]{#2}
\providecommand{\eprint}[2][]{\url{#2}}

\end{thebibliography}


\begin{references}
\bibitem{matsukura}%"review：Control of magnetism by electric fields,"
F. Matsukura, Y. Tokura, and H. Ohno, Nat. Nanotechnol. \textbf{10}, 209(2015).

\bibitem{brataas}%Review：Current-induced torques in magnetic materials.;published online EpubMay (10.1038/nmat3311)
A. Brataas, A. D. Kent, H. Ohno, Nat. Mater. \textbf{11}, 372(2012).

\bibitem{ohno}%"Electric-field control of ferromagnetism,"
H. Ohno, D. Chiba, F. Matsukura, T. Omiya, E. Abe, T. Dietl, Y. Ohno, and K. Ohtani, Nature (London) \textbf{408}, 944(2000).

\bibitem{chiba}%"Electrical manipulation of magnetization reversal in a ferromagnetic semiconductor,"
D. Chiba, M. Yamanouchi, F. Matsukura, and H. Ohno, Science \textbf{301}, 943(2003).

\bibitem{lebeugle}%Electric Field Switching of the Magnetic Anisotropy of a Ferromagnetic Layer Exchange Coupled to the Multiferroic Compound BiFeO3,
D. Lebeugle, A. Mougin, M. Viret, D. Colson, L. Ranno, Phys. Rev. Lett. \textbf{103},257601(2009).

\bibitem{heron}%Electric-Field-Induced Magnetization Reversal in a Ferromagnet-Multiferroic Heterostructure,
J.T. Heron, M. Trassin, K. Ashraf, M. Gajek, Q. He, S.Y. Yang, D.E. Nikonov, Y.H. Chu, S. Salahuddin, R. Ramesh, Phys. Rev. Lett. \textbf{107}, 217202(2011).

\bibitem{tokunaga}%"Electric-field-induced generation and reversal of ferromagnetic moment in ferrites,"
Y. Tokunaga, Y. Taguchi, T.-h. Arima, Y. Tokura, Nat. Phys. \textbf{8}, 838(2012).

\bibitem{weisheit}%"Electric field-induced modification of magnetism in thin-film ferromagnets,"
M. Weisheit, S. Faehler, A. Marty, Y. Souche, C. Poinsignon, D. Givord, Science \textbf{315}, 349(2007).

\bibitem{maruyama}% Large voltage-induced magnetic anisotropy change in a few atomic layers of iron
T. Maruyama, Y. Shiota, T. Nozaki, et al. \Journal{Nat. Nanotechnol.} {4} {158} {2009}.

\bibitem{shiota}%Induction of coherent magnetization switching in a few atomic layers of FeCo using voltage pulses,
Y. Shiota, T. Nozaki, F. Bonell, S. Murakami, T. Shinjo, Y. Suzuki, Nat. Mater. \textbf{11}, 39(2012).

\bibitem{slon1}%Current-driven excitation of magnetic multilayers
J. C. Slonczewski. J. Magn. Magn. Mater., \textbf{159}, L1-L7 (1996); J. Magn. Magn. Mater., \textbf{195}, L261-L268 (1999).

\bibitem{berger}
L. Berger, Phys. Rev. B. \textbf{54}, 9353(1996); J. Appl. Phys., \textbf{90}, 4632 (2001).

\bibitem{tsoiprl} M. Tsoi, A.G.M. Jansen, J. Bass, W.C. Chiang, M. Seck, V. Tsoi and P. Wyder, \Journal{Phys. Rev. Lett.}{80}{4281}{1998}; \textbf{81}, 493(E) (1998).

\bibitem{cornellorig} J.A. Katine, F.J. Albert, R.A. Buhrman, E.B. Myers, and D.C. Ralph, \Journal{Phys. Rev. Lett.}{84}{3149}{2000}.

\bibitem{miron}%"Perpendicular Switching of a Single Ferromagnetic Layer Induced by in-Plane Current Injection."
I. M. Miron, K. Garello, G. Gaudin, P. J. Zermatten, M. V. Costache, S. Auffret, S. Bandiera, B. Rodmacq, A. Schuhl, and P. Gambardella. \Journal{Nature}{476}{189}{2011}.

\bibitem{suzuki}%"Current-Induced Effective Field in Perpendicularly Magnetized Ta/CoFeB/Mgo Wire."
T. Suzuki, S. Fukami, N. Ishiwata, M. Yamanouchi, S. Ikeda, N. Kasai, and H. Ohno. \Journal{Appl. Phys. Lett.} {98}{142505}{2011}.

\bibitem{liulq}%"Current-Induced Switching of Perpendicularly Magnetized Magnetic Layers Using Spin Torque from the Spin Hall Effect."
L. Liu, O. J. Lee, T. J. Gudmundsen, D. C. Ralph, and R. A. Buhrman. \Journal{Phys. Rev. Lett.} {109} {096602} {2012}.

\bibitem{demidov}
V. E. Demidov, S. Urazhdin, H. Ulrichs, V. Tiberkevich, A. Slavin, D. Baither, G. Schmitz, and S. O. Demokritov.
 Nat. Mater. \textbf{11}, 1028 (2012).

\bibitem{diakonov}%Spin Hall effect - Wikipedia, the free encyclopedia
M. I. Dyakonov and V. I. Perel, \Journal{Sov. Phys. JETP Lett}{13}{467}{1971}.

\bibitem{hirsch}% spin Hall effect
J. E. Hirsch, Phys. Rev. Lett. \textbf{83}, 1834 (1999).

\bibitem{rashba}
Y. A. Bychkov and E. I. Rashba, J. Phys. C \textbf{17}, 6039 (1984); G. Dresselhaus, Phys. Rev. \textbf{100}, 580 (1955).

\bibitem{manchon}%"Theory of Spin Torque Due to Spin-Orbit Coupling."
A. Manchon and S. Zhang, \Journal{Phys. Rev. B.} {79}{094422} {2009}.

\bibitem{pi}%"Tilting of the spin orientation induced by Rashba effect in ferromagnetic metal layer,"
U.H. Pi, K.W. Kim, J.Y. Bae, S.C. Lee, Y.J. Cho, K.S. Kim, and S. Seo, Appl. Phys. Lett. \textbf{97}, 162507(2010).

\bibitem{wang}% Electric-field-assisted switching in magnetic tunnel junctions
W. G. Wang, M. Li, S. Hageman and C. L. Chien. \Journal{Nat. Mater.} {11} {64} {2012}.

\bibitem{liul}%Magnetic Oscillations Driven by the Spin Hall Effect in 3-Terminal Magnetic Tunnel Junction Devices,
L. Liu, C.-F. Pai, D.C. Ralph, R.A. Buhrman, Phys. Rev. Lett. \textbf{109}, 186602(2012).

\bibitem{liurh}%Spectral Characteristics of the Microwave Emission by the Spin Hall Nano-Oscillator
R. H. Liu, W. L. Lim, and S. Urazhdin. Phys. Rev. Lett. \textbf{110}, 147601 (2013)

\bibitem{ulrichs} %Optimization of Pt-based spin-Hall-effect spintronic devices
H. Ulrichs, V.E. Demidov, and S.O. Demokritov, W.L. Lim, J. Melander, N. Ebrahim-Zadeh, and S. Urazhdin, \Journal{Appl. Phys. Lett}{102}{132402}{2013}

%\bibitem{nguyen}%Spin Torque Study of the Spin Hall Conductivity and Spin Diffusion Length in Platinum Thin Films with Varying Resistivity
%Minh-Hai Nguyen, D. C. Ralph,and R. A. Buhrman, \Journal{Phys. Rev. Lett.}{116}{126601}{2016}.

\bibitem{slavinprl}%Spin Wave Mode Excited by Spin-Polarized Current in a Magnetic Nanocontact is a Standing Self-Localized Wave Bullet
A. Slavin and V. Tiberkevich, Phys. Rev. Lett. \textbf{95}, 237201 (2005).

\bibitem{kim}%Line Shape Distortion in a Nonlinear Auto-Oscillator Near Generation Threshold: Application to Spin-Torque Nano-Oscillators
Joo-Von Kim, Q. Mistral, C. Chappert, V. Tiberkevich, and A. N. Slavin, Phys. Rev. Lett. \textbf{100}, 167201 (2008).

\bibitem{nonlinear}
A. Slavin and V. Tiberkevich, \Journal {IEEE Trans. Magn.} { 44}{1916}{2008}; \Journal{ibid}{45}{1875}{2009}.

\bibitem{fuchs} G.D. Fuchs, J.C. Sankey, V.S. Pribiag, L. Qian, P.M. Braganca, A.G.F. Garcia, E.M. Ryan, Z.-P. Li, O. Ozatay, D.C. Ralph, and R.A. Buhrman, \Journal{Appl. Phys. Lett.}{91}{062507}{2007}.

\bibitem{urazhdin}%Parametric Excitation of a Magnetic Nanocontact by a Microwave Field
S. Urazhdin, V. Tiberkevich, and A. Slavin, Phys. Rev. Lett. \textbf{105}, 237204(2010).

\bibitem{leix}%Electric field control of interface magnetic anisotropy
L. Xu and S. F. Zhang. \Journal{J. Appl. Phys.} {111} {07C501} {2012}.

\bibitem{liuprb}
R. H. Liu, W. L. Lim, and S. Urazhdin, \Journal {Phys. Rev. B} {89} {220409(R)}{2014}.

\bibitem{rippard}%Current-driven microwave dynamics in magnetic point contacts as a function of applied field angle
W. H. Rippard, M. R. Pufall, S. Kaka, T. J. Silva, and S. E. Russek, \Journal{Phys. Rev. B} {70} {100406(R)} {2004}.

\noindent

\end{references}
\end{document}